# Analogy of space-time as an elastic medium - Study of the perturbation tensor of the metric $h_{\mu\nu}$ through the prism of the analogy of the theory of elasticity- Analysis and potential consequences -

David Izabel[1][3]  Yves Remond[1]  Matteo Luca Ruggiero[2]

[1] University of Strasbourg, CNRS, France
[2] University of Turin, Italy
[3] University of L'Aquila, CNRS, Italy

**Abstract**

A state of the art of the different deformations of space-time measured for more than a hundred years in the case of general relativity in the weak field is carried out. The phenomena of general relativity in low fields targeted are gravitational waves, Lense-thirring effects, gravitational lensing, gravitation around the earth or the sun. The overview of these different deformations highlights the different active components of the perturbation tensor of the metric $h_{\mu\nu}$. The authors show that each phenomenon corresponds to one or more very specific components of this tensor. They also show that the various components of the latter have as an image, within the framework of the elastic analogy of space-time, various coherent components of an associated strain tensor $\varepsilon_{\mu\nu}$ in terms of elongation, shortening or angular distortion of an equivalent elastic medium, modeling the behavior of space-time. By this synthetic ensemble approach and this elastic analogy, it appears clearly for the first time that some components of this tensor $h_{\mu\nu}$ remain to be determined and measured in adequacy with potential new phenomena or modified versions of general relativity in a weak field.

Keywords: General relativity, gravitational waves, frame dragging, geodetic effect, Lense thirring effect, Sun gravitation, Earth gravitation, gravitational lens, elasticity, continuum mechanics

## 1. Introduction

The theory of general relativity [1], [2] is more than a century old and is now tested and validated in many situations with increasingly precise experiments [3], [4], [5], [6], [7], It works well and conceptually gravitation is now associated with the notion of deformation or curvature of space-time. This distortion of space-time is linked to the establishment of the metric tensor, $g_{\mu\nu}$ which is determoned by solving the nonlinear differential tensor equation of the gravitational field established by A. Einstein in 1915 [1].

$$R_{\mu\nu} - \frac{1}{2}g_{\mu\nu}R = \frac{8\pi G}{c^4}T_{\mu\nu} \quad (1)$$

In this expression, $R_{\mu\nu}$ is the Ricci curvature tensor, R is the tensor contraction of the Ricci tensor, $T_{\mu\nu}$ the energy-momentum tensor that somehow shapes space-time, G is Newton's gravitational constant, and c is the speed of light.

In weak gravitational fields, the metric $g_{\mu\nu}$ can be written as:

$$g_{\mu\nu} = \eta_{\mu\nu} + h_{\mu\nu} \quad (2)$$

Where $\eta_{\mu\nu}$ is the Minkovski tensor of flat space-time and $h_{\mu\nu}$ a very small perturbation ($|h_{\mu\nu}| \ll |\eta_{\mu\nu}|$)

The purpose of this article is an in-depth study of the different components of $h_{\mu\nu}$, in the light of the various tests of general relativity carried out over the past 100 years.

Some components of $h_{\mu\nu}$ are activated and their intensities are established according to the physical phenomena soliciting space-time. The phenomena of general relativity in low fields targeted are gravitation around the Earth or the sun, gravitational waves, Lense-Thirring effects (frame dragging, geodetic effect). These effects involve different components of $h_{\mu\nu}$ which we will now detail and study with a synthetic overview. Each of these components are particular deformations of space-time (elongations, shortenings, angular variations) that have been precisely measured via tests carried out in recent years. The objective of this article is therefore to have a global and similar vision of these different components of this tensor $h_{\mu\nu}$ and to understand them in the framework of the elastic analogy with a 4-dimensional elastic strain tensor $\varepsilon_{\mu\nu}$ in order to see if these two tensors actually represent similar mechanical phenomena with identical component numbers on the one hand and if some components of these tensors remain to be studied on the other hand (new physical phenomenon? new polarization in physics associated with degrees of freedom in the framework of elastic medium analogy).

**2. Methods**

The methodology followed by the authors is as follows.

- Identify for each phenomenon of general relativity in weak fields the components of the perturbation tensor $h_{\mu\nu}$ of the metric that are active,

- Synthesize into a global vision its different components simultaneously in the tensor in order to identify the active and non-active components of it,

- Explicitly construct the co-correspondences between the deformations of spacetime associated with their position in the perturbation tensor of the metric $h_{\mu\nu}$ and the deformations relative to an elastic deformation tensor $\varepsilon_{\mu\nu}$ associated with space-time in the case of the analogy to an elastic medium,

- Use the analogy with the deformation tensor of the equivalent elastic medium and the perturbation tensor of the equivalent metric thus established to identify any new physical phenomena or deformations that would remain to be measured,

- Conclude on possible implication on the low energy limit of general relativity to cover these new components of the perturbation tensor of the metric.

## 3 Presentation and structuring of the different components of the perturbation tensor of the metric according to the physical phenomenon of general relativity in the weak field studied

### 3.1 Concerning the activation of the component $h_{00}$ – Newton's classical gravitation

This was developed by Einstein in 1915 **[1].** Einstein's general equation then becomes for the time components, calibrated to find Newton's gravitation via $\kappa = \frac{8\pi G}{c^4}$. In this expression ϕ is Newton's gravitational potential.

$$g_{00} = \eta_{00} + h_{00} = 1 + \frac{2\phi}{c^2} \quad (3)$$

Taking the Laplacian noted Δ from the above expression, we obtain:

$$\Delta h_{00} = \frac{2}{c^2} \Delta\phi \quad (4)$$

Moreover, we know that in a weak field, the gravitational field satisfies the Poisson equation:

$$\Delta\phi = 4\pi G\rho \quad (5)$$

Thus we find:

$$\Delta h_{00} = \frac{2}{c^2} \Delta \phi = \frac{8\pi G}{c^4} \rho c^2 = \frac{8\pi G}{c^4} T_{00} = \kappa T_{00} \quad (6)$$

This equation is the transposition of Einstein's equation into a weak gravitational field. It gives Newton's gravitation, i.e. the Poisson's equation.

Space-time becomes deformable, it becomes a physical object subject to deformation when it is stressed, just like any material elastic medium that can be studied with the theory of elasticity or the resistance of materials (beams, plates, shells). It is therefore in this field of weak gravitational fields and using elastic analogy that we will understand general relativity in this paper.

To visualize this physical object in transparating space-time, since the intergalactic vacuum fills the space around the stars there, an analogy consists in considering a two-dimensional membrane represented in the form of a grid fabric where we can see how the sun and the earth deform it. This membrane approach certainly has its limitations, but it allows us to clearly visualize the effects of gravitation, such as the apparent shift of stars behind the sun. This was shown in the publication **[52]**. Marbles thrown on this surface following the curvature of the canvas will inexorably stick to a petanque ball placed in the center simulating the appearance of a massive object like the sun, giving the impression of a Newtonian force that attracts them towards each other.

But since gravitation is a 3-dimensional phenomenon under Newton and a 4-dimensional phenomenon for Einstein, this analogy necessarily has limit: namely, according to Einstein both space and time are deformed **[60]** In addition, Einstein calibrated his field equation to the temporal component $\boldsymbol{h_{00}}$ of its tensors to find Newton's gravitation, which works particularly well in the solar system in particular **[51]**.

**3.2 Concerning the activation of the components $h_{00} h_{xx}, h_{yy}, h_{zz}$, associated with Newton's classical gravitation applied to gravitational lensing – case on single ray of light**

If a mass energy deforms space very slightly, a gravitational field intervenes and the interval in this weakly disturbed space-time is then written **[8]**:

$$\boldsymbol{ds^2 = -\left(1 + \frac{2\phi}{c^2}\right)c^2 dt^2 + \left(1 - \frac{2\phi}{c^2}\right)(dx^2 + dy^2 + dz^2)} \quad (7)$$

The metric of this space-time is then written from the Minkovsky tensor and perturbation tensor of the metric:

This metric is used in the case of gravitational lensing **[8]**. The interval is of the light type $\boldsymbol{ds^2 = 0}$

In this case according to **[8]**, we can obtain the propagation coordinate time:

$$t = \frac{1}{c}\int_S^0 \left(1 - \frac{2\phi}{c^2}\right)dl = \frac{D}{c} - \frac{2}{c^3}\int_S^0 \phi dl \quad (8)$$

So, gravitational lensing depends both on the Newtonian potential $h_{00}$ and on the space curvature $h_{ii}$.

Thus $h_{\mu\nu}$ can be sinded into 3 parts, the temporal components $h_{00}$, the spatiotemporal components $h_{0i}$ and $h_{j0}$ the purely spatial components $h_{ij}$. For each of these components, a very specific phenomenon of general relativity in a weak field must be considered. We will therefore look at these components in the next paragraph.

### 3.3 Concerning the components $h_{ij}(i, j \to x, y)$ – gravitational waves

This was developed by A. Einstein in 1916 **[2]** and then in 1918 **[9].** The components $h_{ij}$ of the perturbation tensor of the metric are associated with gravitational waves.

In weak-field appr oximation Einstein's field equation becomes **[9]**:

$$\partial^\lambda \partial_\lambda \overline{h}_{\mu\nu} = \Box \, \overline{h}_{\mu\nu} = -\frac{16\pi G}{c^4} T_{\mu\nu} \quad (9)$$

In the specific case of gravitational waves when placed in a vacuum $T_{\mu\nu} = 0$, the above equation becomes:

$$\partial^\lambda \partial_\lambda \overline{h}_{\mu\nu} = \Box \, \overline{h}_{\mu\nu} = 0 \quad (10)$$

h is the trace of $h_{\mu\nu}$, $\Box$: the D'Alembert's operator. the gauge condition taken is $\partial^\lambda \overline{h}_{\mu\lambda} = 0$

$\overline{h}_{\mu\nu}$ is the perturbation of the metric based on the next variable change.

$$\overline{h}_{\mu\nu} = h_{\mu\nu} + \frac{1}{2}\eta_{\mu\nu}\overline{h} \quad (11)$$

The solution of this differential equation is of the form:

$$\overline{h}_{\mu\nu} = -\frac{4G}{c^4}\iiint_{Source} \frac{T_{\mu\nu}(x^0 - x_s^0 - \|\vec{x} - \vec{x}_s\|)}{\|\vec{x} - \vec{x}_s\|} d^3\vec{x}_s \quad (12)$$

In the case of gravitational waves, the solution of the above equation in vacuum can be writren as :

$$\overline{h}_{\mu\nu} = A_{\mu\nu}\cos(k_\sigma k^\sigma) \quad (13)$$

With:

$$A_{\mu\nu} = A_+ \begin{pmatrix} 0 & 0 & 0 & 0 \\ 0 & 1 & 0 & 0 \\ 0 & 0 & -1 & 0 \\ 0 & 0 & 0 & 0 \end{pmatrix} + A_\times \begin{pmatrix} 0 & 0 & 0 & 0 \\ 0 & 0 & 1 & 0 \\ 0 & 1 & 0 & 0 \\ 0 & 0 & 0 & 0 \end{pmatrix} \quad (14)$$

The non-zero terms are therefore at the coordinates $xx, yy, xy, yx$. We therefore note the perturbation of the spatial metric $h_{ij}$ and no longer the spatio-temporal one. Given the gauge conditions, only the space part related to the deformations of these gravitational waves remains. These components are related to the sources of gravitational wave **[9]**, in fact we have:

$$\bar{h}_{ij(t)} = h_{ij(t)} = \frac{2G}{Rc^4} \frac{d^2}{dt^2} I_{ij}\left(t - \frac{R}{c}\right) = \frac{2G}{Rc^4} \ddot{Q}_{ij}\left(t - \frac{R}{c}\right) \quad (15)$$

With R the distance of the observer with respect to the source, $I_{ij}$ quadrupole moment, $Q_{ij(t)}$ the quadripolar moment. The deformations related to gravitational waves are therefore spatial and manifest themselves by elongations and shortening in planes $(x, y)$ (perpendicular to the direction of propagation $z$ in this article) of these waves. For simplification, we will talk about deformations in the plane in this article. Gravitational waves were first measured in 2014 via the GW150914 event **[6]**. Gravitational and electromagnetic waves were measured simultaneously for the first time during the GW170817 event **[7]**.

**3.4 Concerning the off-diagonal components $h_{0i}$ $h_{j0}$ : frame-dragging effect and geodesic precession**

This was developed by J. Lense and H. Thirring in 1918 **[3]**. In this case, it is the effect of the frame of reference being driven by a massive rotating object that is at stake (for example, the rotation of the earth or the sun). In paper **[3]**, formula 4 about $T_{\mu\nu}$ and 10, $g_{\mu\nu}$, we can see that the off diagonal component $h_{0i}; h_{j0}$ caracterise this effect. Recent publications **[61]** these effects are studied also with more recent notation.

The two Lense Thirring effects (frame dragging and geodetic) were measured via the gravity prob B experiment **[5]**.

**4 Summary and overview of the different components of the perturbation tensor of the metric**

In view of the previous chapter 3, in a weak field, general relativity does indeed predict deformations (elongations, shortenings, angles) in the plane and perpendicular to the plane. Figure 1 below illustrates each phenomenon associated with each component of the perturbation tensor of the metric $h_{\mu\nu}$ that we will study in this article ($0 = t, 1 = x; 2 = y; 3 = z$). By superimposing these different components of the tensor $h_{\mu\nu}$ to have for the first time a global and synthetic vision, we can see that some components remain inactive $h_{0z}; h_{iz}; h_{zj}$ (with regard to the classical phenomena measured described in the previous chapter 3).

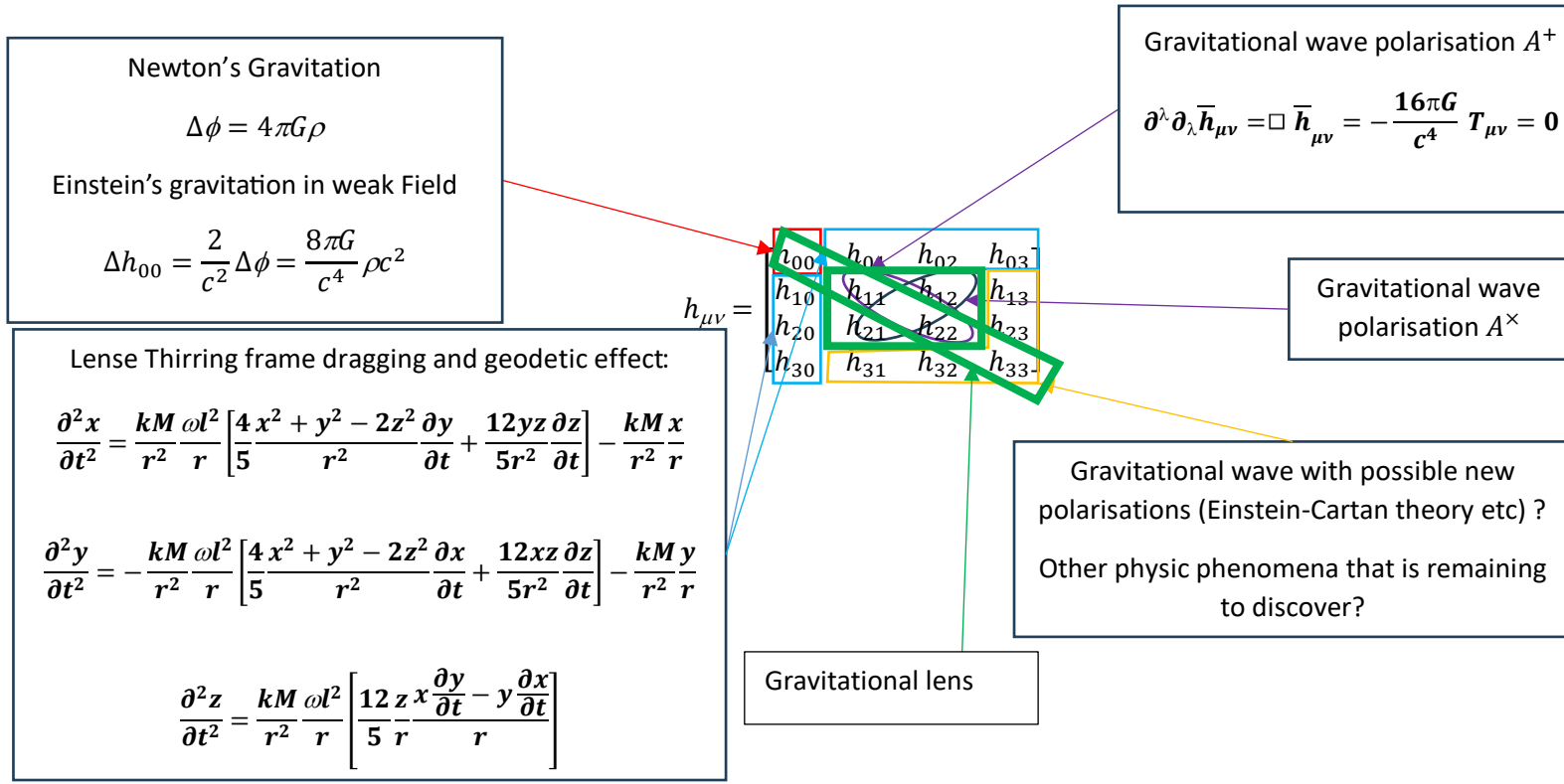

**Figure 1: Physical phenomena associated with each component of the perturbation tensor of the metric -$h_{\mu\nu}$**

## 5. Deep co-correspondence between the deformations of spacetime from the perturbation tensor of the metric and a deformation tensor of an equivalent elastic medium

*5.1 Founding principles and Hooke's law associated with the elastic analogy of space-time*

The basic principles of equivalence are as follows. They come from well-known publications or in fact in the context of the elastic analogy of the space-time environment.

**Equivamence principle n°1: Correspondence between the perturbation tensor of the metric and the strain tensor in the analogy of the elastic medium**

This principle demonstrated in **[10]**, **[11]** and **[12]** is as follows:

$$\boldsymbol{h}_{\mu\nu} = \boldsymbol{2}\varepsilon_{\mu\nu} \quad (16)$$

With:

$h_{\mu\nu}$ the perturbation tensor of the metric such that defines in the metric of the general relativity from the flat Minkovski metric

$\varepsilon_{\mu\nu}$ the strain tensor in 4 dimensions.

Both are symmetric tensors.

**Equivalence principle n°2: Equivalence between the energy-momentum tensor and the stress tensor in elasticity**

This principle demonstrated in **[11]** is as follows:

In four dimensions $T_{\mu\nu} = \rho_{matter} u_u u_v$ . in three dimensions $\sigma_{ij} = \rho v_i v_j$

In these expressions, $T_{\mu\nu}$ is the tensor energy momentum, $u_u$ and $u_v$ the four velocities, $\sigma_{ij}$ the tensor of constraints, $v_i$, and $v_j$ the velocities.

**Equivalence principle n°3: Einstein's constant characterizes the flexibility of space-time**

This principle, demonstrated in **[11]** and **[12],** is as follows:

- For an elastic bar of length L with a cross-section S of Young's modulus Y = E stressed by a normal force N undergoing longitudinal displacements $u_{(x)}$ along the direction $x$ and therefore deformations , $\varepsilon_{xx}$ and $\varepsilon_{yy}$ :

$$\left(\frac{du}{dx}\right)^2 = \frac{2}{ES}\frac{U}{L} = \frac{2}{YS}\frac{U}{L} \ (17)$$

$$U = \frac{1}{2}\int_0^L \frac{N^2}{ES} dx \ (18)$$

- For a elastic bar of length L with a reduced shear modulus section $S_r$ of shear modulus G = μ stressed by a shear force V undergoing transverse displacements along the direction $yx$ and therefore deformations $\varepsilon_{xy}$, and $\varepsilon_{yx}$ associated with angular distortion $\gamma_{xy}$, and $\gamma_{yx}$ :

$$\left(\frac{dy}{dx}\right)^2 = \frac{2}{GS_r}\frac{U}{L} = \frac{2}{\mu S_r}\frac{U}{L} \ (19)$$

$$U = \frac{1}{2}\int_0^L \frac{V^2}{GS_r} dx \quad (20)$$

These flexibilities $2/ES$ and $2/GS_r$ in one wireframe dimension are compatible (same unit N⁻¹) with the flexibility constant $\kappa = \frac{8\pi G}{c^4}$ (with G the gravitational constant) in 4 dimensions in Einstein's gravity field equation. Let us generalize for space alone with Hooke's law according to **[10]**:

$$U_\xi = \frac{1}{2}\sigma^{ij}\varepsilon_{ij} = \frac{1}{2}C^{ijkl}\varepsilon_{ij}\varepsilon_{kl} \quad (21)$$

With:

$$C^{ijkl} = \frac{Y}{1+\nu}\left(\frac{\nu}{1-2\nu}g^{ij}g^{kl} + g^{ik}g^{jl}\right) \quad (22)$$

**Equivalence principle n°4: In a field of weak gravity, the Einstein field equation is equivalent to a Hooke's law**

According to **[10]**, **[11]**, **[12]**, **[14]** in a weak field Einstein's equation is written as defined in 9.

In the case of the transverse gauge, the trace of $h_{\mu\nu}$ h = $\bar{h}$ = 0 and therefore $\bar{h}_{\mu\nu} = h_{\mu\nu}$ = 2 $\varepsilon_{\mu\nu}$ according to principle 1

$T_{\mu\nu}$ is related to the stress tensor according to principle 2

$\frac{16\pi G}{c^4} = 2\kappa$ and is related to the flexibility of the equivalent elastic medium via 1/E=1/Y according to principle 3.

Everything is therefore conceptually equivalent with a Hooke's law of an equivalent elastic medium that behaves like spacetime:

$$\varepsilon = \frac{1}{E}\sigma \quad (23)$$

$$\gamma = \frac{1}{G}\tau \quad (24)$$

**Equivalence principle n°5: The Young's modulus of the elastic medium can be associated with the energy density of the medium**

In the case of elastic waves, it is shown in **[10]** and **[11]** and **[12]** that:

- In the case of longitudinal waves:

$$Y = \rho_{vacuum}c^2 \quad (25)$$

- In the case of shear waves:

$$\mu = \rho_{vacuum} c_{shear}^2 \quad (26)$$

With ρ the density of the medium, the vacuum for the purposes of this article.

**Principle n°6: The polarization of gravitational waves can be considered as components associated with a strain tensor in a vacuum modeled as an elastic medium:**

This principle demonstrated in **[11]** is based on principle 1 and is as follows:

$$h_{\mu\nu} = A_+ \cos\left(\frac{\omega}{c}(ct-z)\right) \begin{bmatrix} 0 & 0 & 0 & 0 \\ 0 & +1 & 0 & 0 \\ 0 & 0 & -1 & 0 \\ 0 & 0 & 0 & 0 \end{bmatrix} (48) \rightarrow \varepsilon_{xy\,(A_+)} = \frac{1}{2} A_+ \cos\left(\frac{\omega}{c}(ct-z)\right) \begin{bmatrix} \varepsilon_{xx} & 0 & 0 \\ 0 & -\varepsilon_{yy} & 0 \\ 0 & 0 & 0 \end{bmatrix}$$

$$h_{\mu\nu} = A_\times \cos\left(\frac{\omega}{c}(ct-z)\right) \begin{bmatrix} 0 & 0 & 0 & 0 \\ 0 & 0 & +1 & 0 \\ 0 & +1 & 0 & 0 \\ 0 & 0 & 0 & 0 \end{bmatrix} \rightarrow \varepsilon_{xy\,(A_\times)} = \frac{1}{2} A_\times \cos\left(\frac{\omega}{c}(ct-z)\right) \begin{bmatrix} 0 & \varepsilon_{xy} & 0 \\ \varepsilon_{yx} & 0 & 0 \\ 0 & 0 & 0 \end{bmatrix} (27)$$

**Postulate n°1 Necessity of considering in a vacuum an energy of deformation of space-time itself**

Classical general relativity gives the response of spacetime when it is charged within it by masses/energies.

If we look at the deformation of spacetime in a vacuum, the classical energy-momentum tensor $T_{\mu\nu}$ becomes zero. In order to retain a Hooke's law associating deformations and stresses via a constant of proportionality, one solution among other, is for example to complete it with a complementary tensor related to the elastic strain energy of the vacuum $t_{\mu\nu,el}$. This is what is proposed in the publications **[49]** and **[50]**. In this case, the Einstein equation becomes:

$$R_{\mu\nu} - \frac{1}{2} R g_{\mu\nu} = \frac{8\pi G}{c^4}\left(T_{\mu\nu} + t_{\mu\nu}\right) \quad (28)$$

In the void:

$$T_{\mu\nu} = 0 + t_{\mu\nu,elastic\ of\ vacuum} \quad (29)$$

A similar approach is followed in when we want to establish the energy of gravitational waves in a vacuum. The Einstein field equation becomes:

$$G_{\mu\nu}^{(1)} = -\frac{8\pi G}{c^4}\left(T_{\mu\nu} + t_{\mu\nu}\right) \quad (30)$$

$G^{(1)}_{\mu\nu}$ is constructed from terms of $G_{\mu\nu}$ which are linear in $h_{\mu\nu}$ (see formula 4 of [53]):

$$t_{\mu\nu} = T^{GW}_{\mu\nu} = \frac{c^4}{8\pi G}\left[G^{(2)}_{\mu\nu} + \cdots\right] \quad (31)$$

$G^{(2)}_{\mu\nu}$ is constructed from quadratic terms of $h_{\mu\nu}$ (see formula 5 of [53]):

*5.2 The mechanical constitution of the equivalent elastic medium behaving like space-time*

The literature to define this equivalent elastic medium is quite dense. We can cite the works of Tenev and Horstemeyer [10], Izabel [11], to [13], and many others [14] to [21].

Generally speaking, the properties that stand out are the following:

A Young's modulus of the order of $10^{113}$ Pa [10] to [13] and [15] obtained from the energy of the vacuum in quantum field theory and of the order of $10^{31}$ Pa ($10^{20} Y_{steel}$) from the elastic energy of the deformations caused by gravitational waves [22].

The Poisson's ratio is 1 in the literature [10], [11] to [13] when analyzing the typical deformations of gravitational waves. A "particle size" of the medium tending towards Planck's length [10], [11] to [13], [15], [32].

A more in-depth analysis from the point of view of the mechanics of gravitational waves [21] tends to show an anisotropic behavior of the elastic medium (deformations of the space of the planes perpendicular to the direction of the gravitational waves and no deformation in the direction of propagation of these waves). All this happens under the dynamic behavior of gravitational waves as if space and the associated elastic medium were made up of leaves without coherence between them.

If we add the geometric torsion in addition to the Riemann curvature tensor, we obtain a modified general relativity which in this case generates complementary polarizations and, under principle 6, complementary deformations between the space sheets [22] to [24]. This phenomenon also appears in gravito-electromagnetism by considering general relativity in the second order [25] [26] or if we consider space-time as a fluid in hydrodynamics and light as sound waves [27]. We thus reconstitute a coherent elastic medium in 3 dimensions. It should be noted that geometric torsion presents a mathematical formalism similar to that of defect theory and plastic crystallography [22] to [26].

On the basis of this state of the art, we can therefore consider that space-time can be modeled by an elastic medium which, in weak fields, follows a Hooke's law but which has characteristics of very weak (almost perfectly rigid) flexibility.

*5.3 Systematic correspondences between the nature of established and measured weak field space-time deformations and deformations of an equivalent strain tensor*

It is well known in elasticity theory that the diagonal components of the strain tensor are related to elongations and shortening and that the other components are related to angular deformations.

If we superimpose the different components of the tensor $\boldsymbol{h}_{\mu\nu}$ with the tensor $\varepsilon_{\mu\nu}$, we see that:

$\boldsymbol{h}_{00}$ (associated with cxt), the first term of the diagonal related to the tensor time which corresponds to isotropic compression in all directions, the classical gravitation that assembles the Earth, the planets, the stars and other stars,

$\boldsymbol{h}_{ii} = -\boldsymbol{h}_{jj}$ diagonal terms related to the tensor space associated with polararion $\boldsymbol{A}^+$, which correspond to the spatial elongations and shortening measured by Ligo, Virgo type interferometers during the passage of a gravitational wave,

$\boldsymbol{h}_{ij} = \boldsymbol{h}_{ji}$ transverse diagonal terms related to space, associated with the polararion $\boldsymbol{A}^\times$, which correspond in torsion according to the orientation of the facets (45°) associated with angular distortions that remain to be measured by the interferometers during the passage of a gravitational wave,

$\boldsymbol{h}_{0i}; \boldsymbol{h}_{j0}$ term of rows and columns associated with time that correspond to the angles generated in the plane and perpendicular to the orbital plane of a rotating star by the Lense-Thirring effect, respectively frame dragging and geodetic effect.

An important point. We consider here different components of the metric perturbation, but as they are adapted to different frames of reference (the center of the Earth for the LT effect, the coordinates of the TT gauge for gravitational waves), they cannot be put directly to compare them in the same tensor. it should first have to be write in the same referential. Nevertheless, the main message here is that the activated components in a general $\boldsymbol{h}_{\mu\nu}$ tensor written in the same coherent framework, which should cover all phenomena simultaneously, will be identified as specified above and could be put in parallel, with the mechanical deformations. by analogy with a 4-dimensional deformation tensor of the space-time elastic medium.

**For physic and general relativity point of view**

$$\begin{bmatrix} h_{tt} & h_{tx} & h_{ty} & h_{tz} \\ h_{xt} & h_{xx} & h_{xy} & h_{xz} \\ h_{yt} & h_{yx} & h_{yy} & h_{yz} \\ h_{zt} & h_{zx} & h_{zy} & h_{zz} \end{bmatrix} \to 2 \times \begin{bmatrix} \varepsilon_{tt} & \frac{1}{2}\gamma_{tx} & \frac{1}{2}\gamma_{ty} & \frac{1}{2}\gamma_{tz} \\ \frac{1}{2}\gamma_{xt} & \varepsilon_{xx} & \frac{1}{2}\gamma_{xy} & \frac{1}{2}\gamma_{xz} \\ \frac{1}{2}\gamma_{yt} & \frac{1}{2}\gamma_{yx} & \varepsilon_{yy} & \frac{1}{2}\gamma_{yz} \\ \frac{1}{2}\gamma_{zt} & \frac{1}{2}\gamma_{zx} & \frac{1}{2}\gamma_{zy} & \varepsilon_{zz} \end{bmatrix} \quad (32)$$

**For physic and mechanical analogy**

## 6. Consequences of analogy - identifications of components not yet associated with physical phenomena in general relativity and possible new deformations, degree of freedom of space, time and polarization

As we have shown in the publication **[11]** the components $\boldsymbol{h_{ij}}$, by analogy with the presumed deformation tensor of distortion associated with a torsion of space twisted by the rotation of two stars rotating around each other, i.e. angular movements in their planes of the interferometer arms that remain to be measured. This should be done via the new generation of interferometers like Lisa.

The components $\boldsymbol{h_{0z}, h_{xz}, h_{yz}}$ do not correspond to deformations identified and measured in classical general relativity.

It is also important to remember that the degrees of freedom associated with the deformations associated with gravitational waves correspond in physics to the polarizations of these same waves. Thus, if the $\boldsymbol{h_{0z}, h_{xz}, h_{yz}, h_{zz}}$ are attached to complementary polarizations, this will result in complementary degrees of freedom in mechanics and components of angular deformations (for $\boldsymbol{h_{0z}, h_{xz}, h_{yz}, h_{zz}}$ ) and elongation/shortening for in the direction of propagation of these waves, as we have shown in **[21].**

These complementary polarizations have not yet been measured and therefore remain speculative at this stage. It should be remembered that they do not appear in the framework of classical general relativity but only in the framework of modified general relativity such as Einstein-Cartan or others.

## *7. Possible experimental validation - optimization of gravitational wave measurements to detect other polarizations associated with other strain tensor deformations and the perturbation tensor of the metric*

In order to complete the components of the perturbation tensor of the metric and in particular $h_{0z}, h_{xz}, h_{yz}$, and their symmetric, several avenues are possible.

- Complementary polarizations in the direction of propagation of gravitational waves not yet measured **[10]** to **[13]**, **[21].**

- Hypothetical stresses of pure shear of space.

In any case, tensors are always one step ahead of physicists, as P Langevin said, so these boxes of components cannot remain empty. Certainly, other phenomena not yet explored by classical general relativity or modified remain to be discovered and will allow them to be completed.

LISA interferometers, the Einstein telescope and the use of pulsars are planned to detect these possible new polarizations or complementary distortions of space-time.

They should also be able to measure possible lateral angular deviations of the interferometer arms **[11]**, **[21].**

## 8. Conclusion

We show that the simultaneous analysis of the different components of the perturbation tensor of the metric, $h_{\mu\nu}$ which has as a mirror in the elastic analogy of gravitation twice the elastic strain tensor $\varepsilon_{\mu\nu}$, allows to better understand general relativity through the prism of the mechanics of continuous media. The components $h_{00}$, $h_{0i}$, $h_{j0}$, $h_{ij}(x,y)$ make it possible to cover physical measurements (elongations/shortening or angular strains) that are rigorously compatible by mirror with that of a strain tensor equivalent to space-time in a weak field that effectively behaves like an elastic medium, if of course all the components of $h_{\mu\nu}$ associated at each physcal phenomena are corrected to be expressed in the same referential. Einstein's field equation appears to be equivalent to a generalized Hooke's law in 4 equivalent spatial dimensions since in the interval $ds^2$ time is associated with the speed of light, thus reconstructing a 4th dimension of length squared $c^2dt^2$. The speed of light is what it is since it is the square root of the ratio between the Young's modulus of the tissue of space-time and the density of the equivalent medium. Since space-time is a dynamic object, it is not possible to have instantaneous absolute deformations. Some of the deformations at a given point are in a way arriving. The polarizations of gravitational waves in classical general relativity are 2 in number, as are the two expressions of the strain tensor in pure torsion following the 45° facets of each other that are considered. This would potentially result in complementary lateral movements of the interferometer arms that are not measurable on LIGO/VIRGO but potentially measurable on the Einstein or Lisa telescopes. The special relativity interval is an elastic strain equation integrating these two aspects in the arrival of deformations at a given point and time. In a way, we measure spatial deformations at a given point, minus those that have not yet arrived and that observers at different points perceive and therefore quantify in different ways. Space-time has mechanical characteristics that have nothing to do with those of conventional materials on earth. Young's modulus differs depending on whether we consider deformations in the plane or perpendicular to the plane of propagation. Orders of magnitude of the order of $10^{20}$ Pa for deformations perpendicular to the plane and $10^{40}$ Pa in the plane **[12]**, **[13]**. A Poisson's ratio of 1 in the plane. The energy engine of time therefore does not seem to be the same as that of space if we use the correlation between Young's modulus and the energy density of the medium.

This analysis of the perturbation tensor of the weak field metric on known and documented experiments of general relativity (gravitational waves GW150914, GW170817, Gravity prob B, terrestrial or solar gravitation) also shows that the elastic analogy of gravitation has a predictive role since it allows us to consider other deformations of space-time, very small for which general relativity in its classical form without geometric torsion passes to side. First of all, complementary polarizations seem to have to be added to the two classics of general relativity (components $h_{0z}, h_{xz}, h_{yz}, h_{zz}$ and their sysmetrics. This results in a complement of the components of the strain tensor to the pure compression cords, shear and angular distortion.

The deformations of space-time in a weak field characterized by the perturbation tensor of the metric $h_{\mu\nu}$ therefore behave, according to our analogy, as an elastic medium characterized by a deformation tensor in 4 dimensions $\varepsilon_{\mu\nu}$ grouping different types of elongations and shortening along the diagonal of the tensor and angular deformations outside the diagonal. Some of its components have already been measured and verified (gravitation $h_{00}$), gravitational wave ($h_{ij}$), entrainment effect of the frame of reference ($h_{0i}$, $h_{j0}$) others remain to be discovered, in particular related to the direction of propagation of gravitational waves $h_{0z}, h_{xz}, h_{yz}, h_{zz}$, such is the message of this article, at least if we believe in the reliability of the advanced elastic analogy. The elastic analogy thus confirms via the co-correspondence between $h_{\mu\nu} = 2\varepsilon_{\mu\nu}$ the various ongoing researches aimed at having other polarizations of space-time in advanced theories of general relativity **[54]** to **[59].**